# Topological Generality and Spectral Dimensionality in the Earth Mineral Dust Source Investigation (EMIT) using Joint Characterization and the Spectral Mixture Residual

D. Sousa[1] and C. Small[2]

[1]*Department of Geography*
*San Diego State University*
*San Diego, CA 92182*
*dan.sousa@sdsu.edu*

[2]*Lamont Doherty Earth Observatory*
*Columbia University*
*Palisades, NY 10984*
*csmall@columbia.edu*

## Abstract

NASA's Earth Surface Mineral Dust Source Investigation (EMIT) mission seeks to use spaceborne imaging spectroscopy (hyperspectral imaging) to map the mineralogy of arid dust source regions. Here we apply recent developments in Joint Characterization (JC) and the spectral Mixture Residual (MR) to explore the information content of data from this novel mission. Specifically, for a mosaic of 20 spectrally diverse scenes we find: 1) a generalized three-endmember (Substrate, Vegetation, Dark; SVD) spectral mixture model is capable of capturing the preponderance (99% in 3 dimensions) of spectral variance with low misfit (99% pixels with <3.7% RMSE); 2) manifold learning (UMAP) is capable of identifying spatially coherent, physically interpretable clustering relationships in the spectral feature space; 3) UMAP yields results that are at least as informative when applied to the MR as when applied to raw reflectance; 4) SVD fraction information usefully contextualizes UMAP clustering relationships, and vice-versa (JC); and 5) when EMIT data are convolved to spectral response functions of multispectral instruments (Sentinel-2, Landsat 8/9, Planet SuperDove), SVD fractions correlate strongly across sensors but UMAP clustering relationships for the EMIT hyperspectral feature space are far more informative than for simulated multispectral sensors. Implications are discussed for both the utility of EMIT data in the near-term, and for the potential of high SNR spaceborne imaging spectroscopy more generally, to transform the future of optical remote sensing in the years and decades to come.

## Keywords

## Introduction

NASA's Earth Mineral Dust Source Investigation (EMIT) mission is designed to study the mineralogy of Earth's dust-forming regions using spaceborne imaging spectroscopy [1]. The EMIT instrument is a Dyson imaging spectrometer with an 11° cross-track field of view, with a fast (F1/8) and wide-swath (1240 samples) optical system achieving roughly 7.4 nm spectral sampling across the 380 – 2500 nm spectral range at high signal-to-noise [2]. EMIT was launched on July 14, 2022 via SpaceX Dragon and successfully autonomously docked to the forward-facing port of the International Space Station (ISS) [3]. EMIT data and algorithms are freely available for public use.

While the stated purpose of the EMIT mission is to measure surface mineralogy and mineral dust in Earth's dust forming regions, these data also provide an unprecedented opportunity to advance our fundamental understanding of the spectral properties of the Earth surface more generally. Sensors like Landsat have collected multispectral satellite imagery for decades [4], but spaceborne hyperspectral (imaging spectroscopy) observations have been much more limited. Early missions like Hyperion [5] and HICO [6] were characterized by nontrivial noise limitations. High quality airborne data from sensors like AVIRIS [7] are available, but with spatial and temporal coverage limitations inherent to airborne platforms. A new generation of spaceborne imaging spectrometers is now starting to come online, with significant involvement from multiple space agencies. Such missions include the Italian Space Agency's CHIME [8] and PRISMA [9], DLR's DESIS [10] and EnMAP [11], JAXA's HISUI [12], and more. EMIT contributes an exciting new aspect to this international constellation and marks an important step towards a global hyperspectral monitoring system.

EMIT began collecting high quality data shortly after launch, and some scenes are already available for download. While geographic coverage of the EMIT mission is inherently limited by the orbital parameters of the ISS, the scenes that have been acquired to date do include broad spectral diversity spanning a wide range of biogeophysical settings. These novel data offer an unprecedented opportunity to test the utility of recent developments in spectral image analysis, including both characterization and modeling.

Here, we use a compilation of 20 spectrally diverse EMIT scenes to investigate the differences between high SNR spaceborne hyperspectral data and simulated multispectral data from common multispectral sensors like Sentinel-2, Landsat, and SuperDove. Specifically, we apply two novel approaches to hyperspectral image analysis: joint characterization [13] and the spectral mixture residual [14]. Joint characterization assumes that important spectral signals may be distributed across multiple scales of variance and provides a way to characterize these signals in a physically interpretable way. The mixture residual uses spectral mixture analysis to isolate low-variance spectral signals (e.g., narrow mineral absorptions) from high-variance signals (e.g., land cover modulated variability in continuum shape and amplitude). Synthesizing these two novel approaches using a novel dataset, we address the following questions:

1) To what extent are EMIT reflectance spectra well characterized by a generalized three-endmember Substrate, Vegetation, Dark (SVD) model, such as has been shown effective for analysis of multispectral satellite imagery?

2) What quantitative and qualitative differences in spectral dimensionality and feature space topology are observed between EMIT reflectance and simulated multispectral data?

3) Does the spectral mixture model residual from EMIT data contain substantially more information than the mixture residual computed from multispectral data? If so, is this effectively captured by traditional dimensionality metrics like variance partition? Does the spectral feature space also manifest as substantially different manifold structure?

4) To what extent can joint characterization be used to reveal subtle but physically meaningful spectral signals in EMIT data? Are these signals spatially coherent?

## Data

20 EMIT scenes were selected on the basis of geographic and spectral diversity (Figure 1). Scenes span 4 continents, sampling important global deserts (Sahara, Arabia, Atacama, Taklamakan, Gobi, Great Basin, Caspian), geologic structure (Zagros, Jabal Tuwaiq, Bushveld, Atacama, Hindu Kush) agriculture (San Joaquin, South African Cape, Hindu Kush), natural floristic diversity (Mata Atlântica, South African Cape, Okavango, Sierra Nevada), as well as some cryospheric targets (Patagonia, Tian Shan) and human settlements (Los Angeles). While this compilation does not achieve comprehensive global sampling, at least some representation is included from most major biomes.

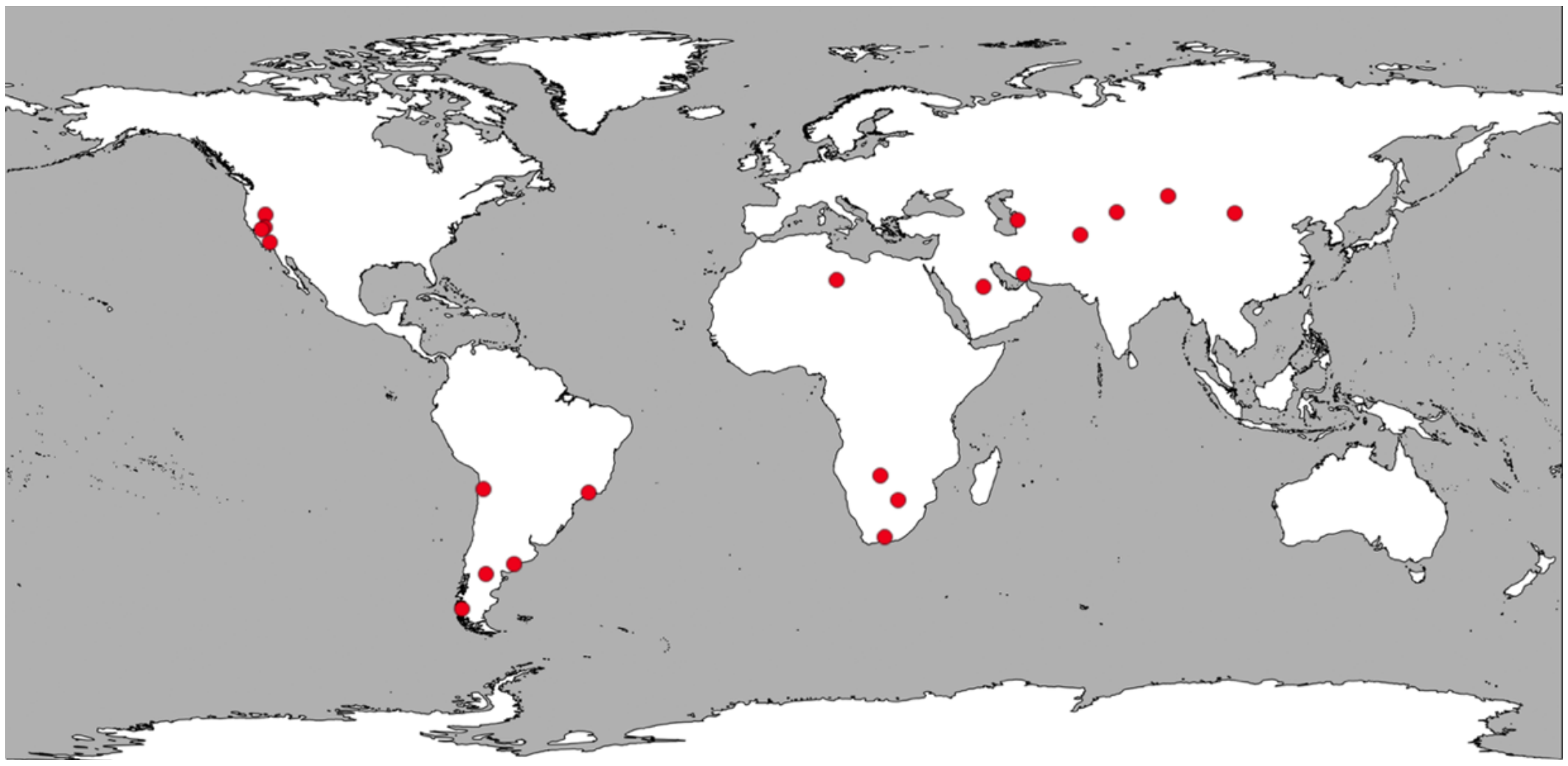

*Figure 1. Index Map. Each of the 20 scenes used for this analysis is shown as a red dot. While the sample is not global, it does include a wide range of land cover including globally significant hotspots for agriculture (San Joaquin), cryosphere (Patagonia, Tian Shan), desert (Taklamakan, Arabian, Saharan, Gobi) and floristic diversity (South African Cape, Mata Atlântica).*

Data were downloaded in netCDF format from the USGS Land Processes Distributed Active Archive Center (LPDAAC) through the web portal: https://search.earthdata.nasa.gov/. Both reflectance and mask files were acquired. All 20 scenes were compiled into a single image mosaic (Figure 2). For subsequent analysis, reflectance data were masked using the "Aggregate Flag" included in Band 7. SceneIDs are provided in Supplementary Table S1.

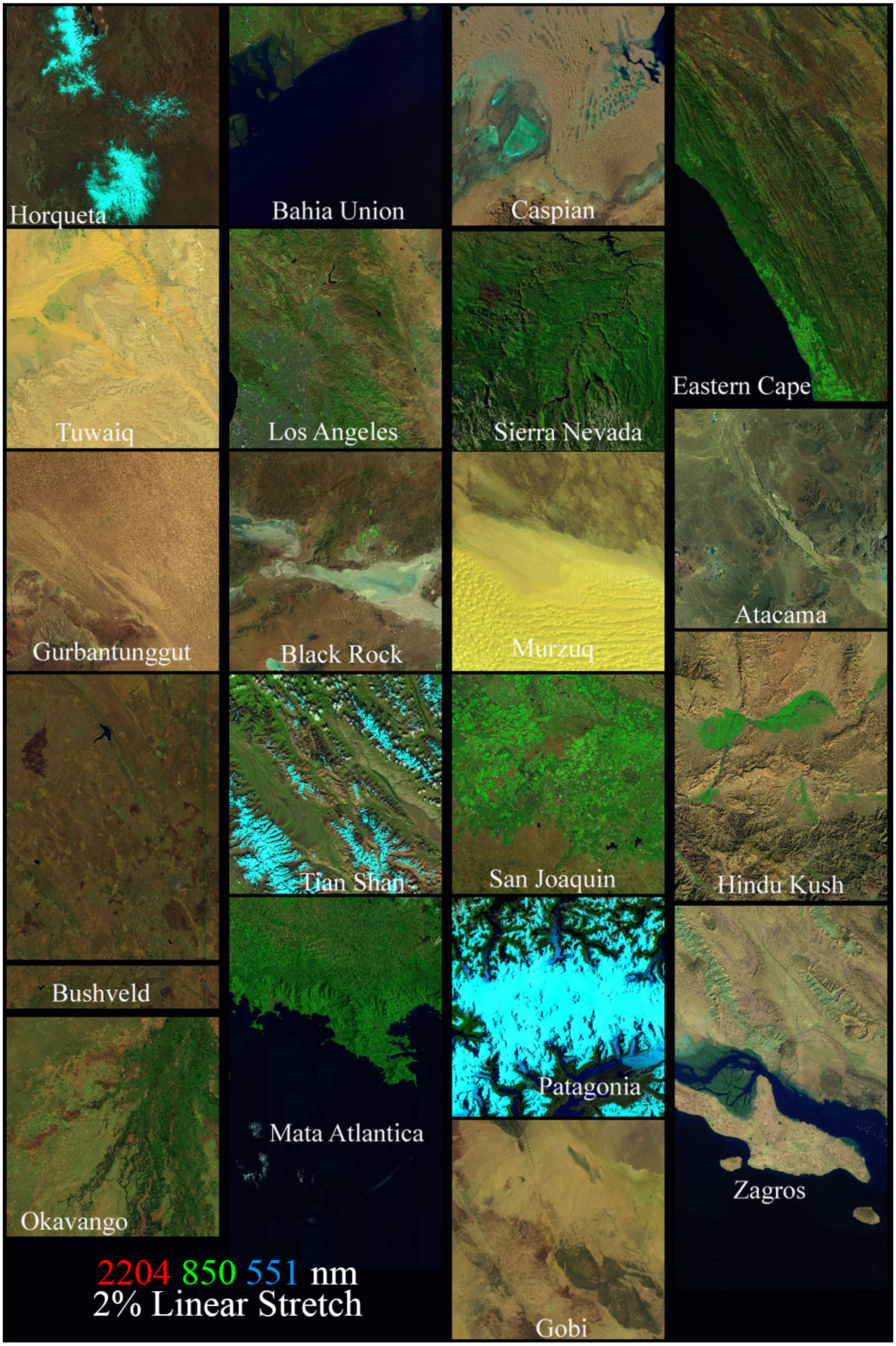

*Figure 2. Mosaic of 20 spectrally diverse EMIT scenes.*

# Analysis

## Variance-Based Spectral Feature Space

Figure 3 shows the low-order variance-based spectral feature space of the reflectance mosaic. The first three dimensions of these data are bounded by snow/ice (I), soil and rock substrates (S), illuminated photosynthetic vegetation (V), and dark targets like shadow and water (D). This low-order topology is consistent with previous regional compilations of AVIRIS imaging spectroscopy [14,15], as well as global compilations of Landsat [16–18], MODIS [19], and Sentinel-2 [20,21]. Significant spectral diversity in the S endmember is observed, associated with geologic variability in sand, bedrock, and soil of the sparsely vegetated scenes in the mosaic. Reflectance spectra for generalized endmembers are shown in the lower right.

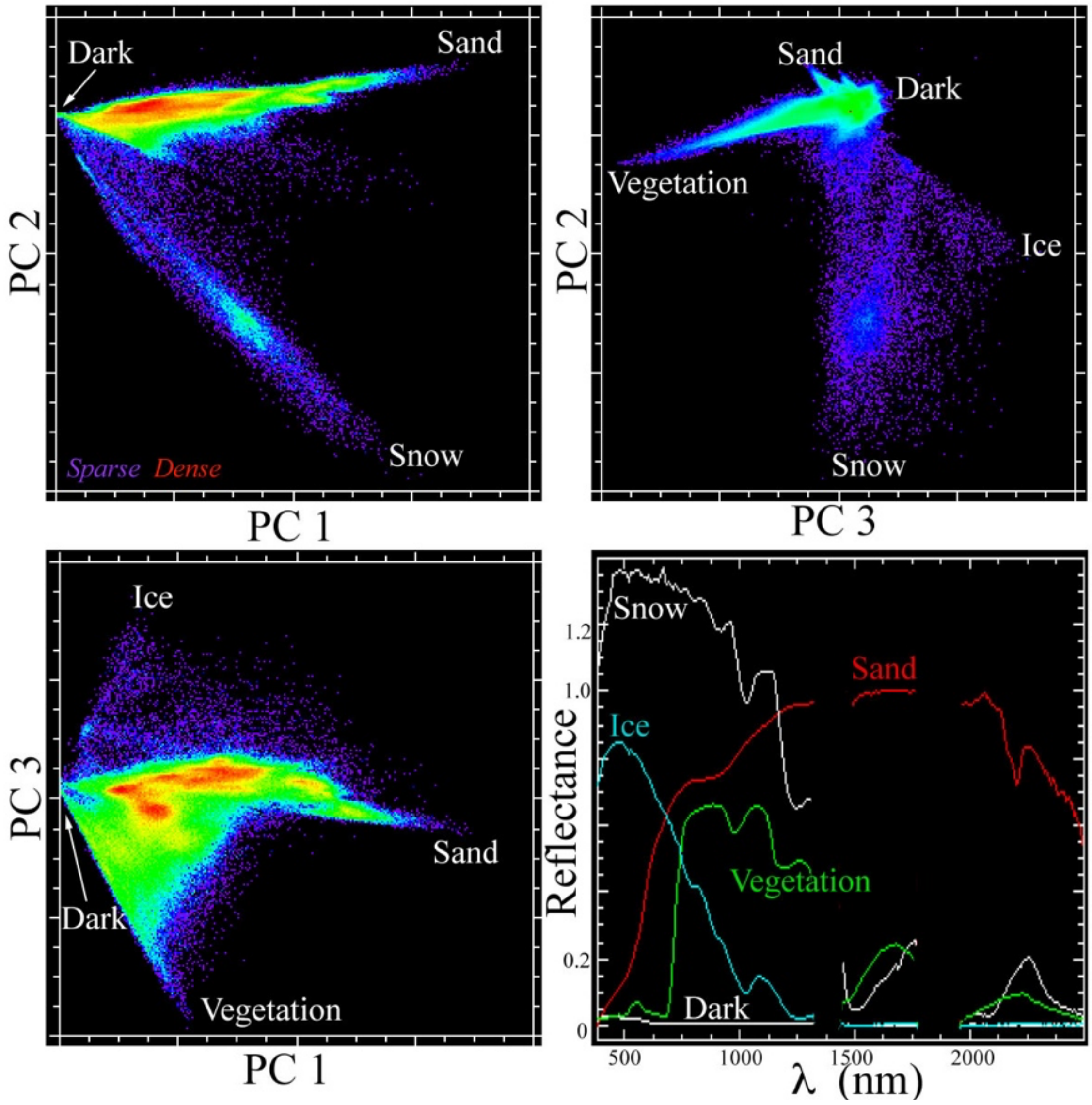

*Figure 3. 3D spectral feature space and spectral endmembers. Scatterplots of orthogonal principal components reveal the straight edges and well-defined apexes of the spectral feature*

*space. Density clustering along the substrate limb between dark and sand results from geologic diversity of arid environments. However, no clusters are geographically specific. All are represented in multiple sample locations.*

The image mosaic was then unmixed using the S, V and D endmembers and wavelength-specific mixture residual was retained, following [14]. The SVD model was found to yield a good fit, with average root-mean-square error (RMSE) approximately 3.1%, and 99% of pixels showing RMSE < 3.7%. The low-order feature space of the mixture residual mosaic is shown in Figure 4. The mixture residual effectively accentuates substrate EM variability by removing the high-variance component of spectral variability which is modeled by a simple linear mixing model. Multiple additional substrate EMs are clearly identifiable in the variance-based mixture residual feature space. Importantly, this endmember variability demonstrates correspondence to VNIR spectral curvature and narrow SWIR absorptions, rather than simple differences in albedo.

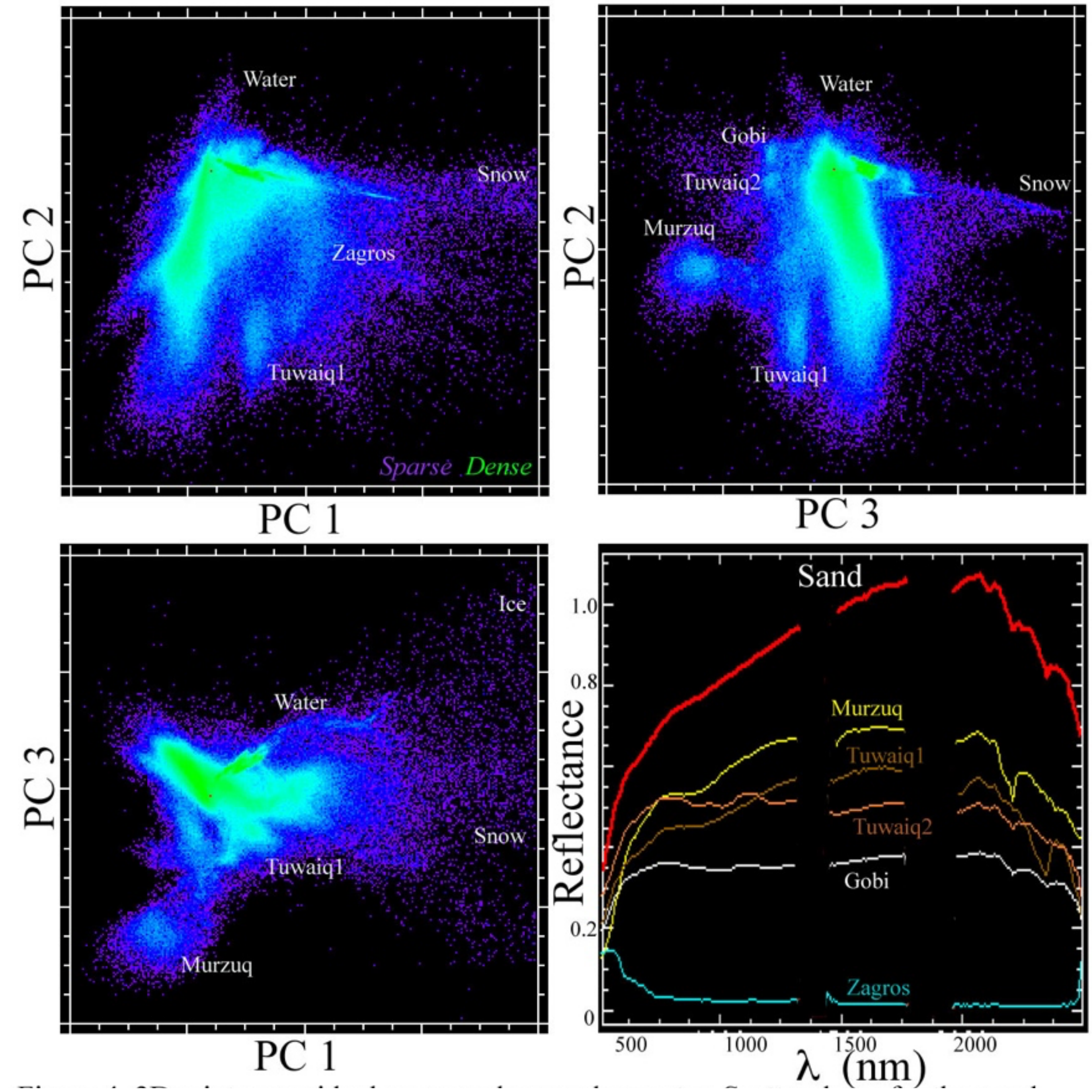

*Figure 4. 3D mixture residual feature space and example spectra. Scatterplots of orthogonal principal components show some geographically distinct clusters (labeled) on periphery but the*

*body of the distribution comingles residuals from almost all sample locations. Tuwaiq 1 and 2 correspond to bedrock and sand respectively. The Zagros spectrum corresponds to shallow water in evaporite pans. The composite sand endmember used to compute the mixture residual is shown in red for comparison.*

In order to investigate EMIT's information content relative to multispectral imagery, the reflectance mosaic was convolved using the spectral response functions of the Sentinel-2A, Landsat 8/9 OLI, and Planet SuperDove sensors. Variance-based characterization was then repeated, including computation of the mixture residual. The difference in information content was then quantified using the partition of variance captured by eigenvalues of the low-order PC dimensions of the reflectance and residual spectra from each sensor (Figure 5). Cumulative variance for the reflectance data (left) shows surprisingly little difference for EMIT data relative to Landsat and Sentinel-2, with minor differences persisting at Dimension 2 but near convergence by Dimension 3. SuperDove shows considerably lower dimensionality, presumably due to the absence of SWIR bands. In contrast, partition of variance from mixture residual spectra (right) shows EMIT dimensionality to consistently exceed all multispectral sensors, and Landsat/Sentinel consistently exceeding SuperDove. The multispectral feature spaces are effectively 2D and 3D while EMIT's hyperspectral feature space is at least 14D to 99.9% of variance.

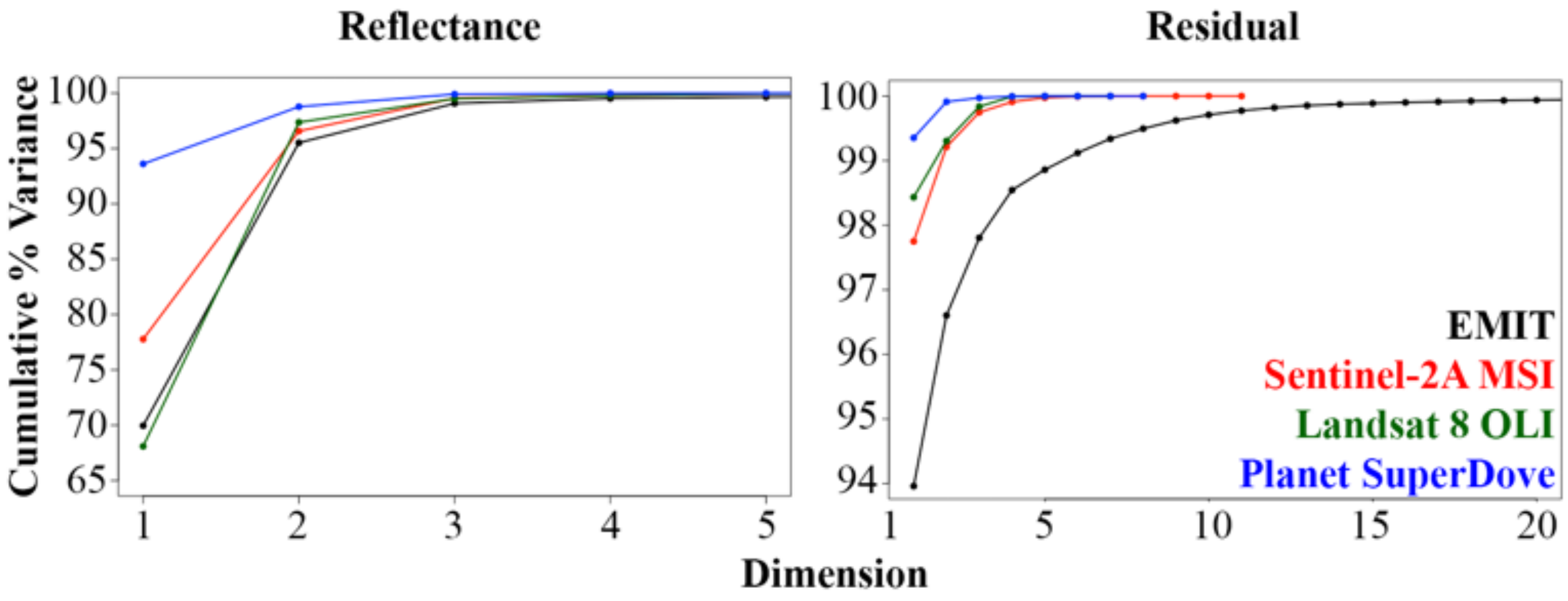

*Figure 5. Partition of variance. When computed from surface reflectance, all 4 sensors show >99% of spectral variance contained in the first 3 dimensions. After computing and removing the generalized (SVD) spectral mixture model, partition of variance much more clearly reflects spectral differences among sensors. EMIT data show highest dimensionality, with 6 additional dimensions required to capture 99% of the remaining variance. Sentinel-2 and Landsat 8 are comparable, each reaching 99% of variance with 2 additional dimensions. SuperDove dimensionality is demonstrably lower, presumably as a result of the lack of SWIR bands. Note differences in both x and y axis scaling between plots.*

## Manifold-Based Feature Space – UMAP

Low-variance spectral feature spaces were further examined using manifold learning. Here, we use the Uniform Manifold Approximation and Projection (UMAP) algorithm [22], implemented using the Python-based 'umap-learn' package. Briefly, UMAP is a nonlinear dimensionality reduction algorithm that assumes the data are uniformly distributed on a locally connected Riemannian manifold with (approximately) locally constant metric. Following these assumptions, the data are modeled using a fuzzy topological structure, and then an embedding is found which maximally approximates (preserves) this topological structure. Both 2D and 3D UMAP embeddings were computed for the mosaic and individual EMIT scenes.

UMAP results for the EMIT mosaic, as well as convolved Sentinel, Landsat, and SuperDove mosaics, are shown in Figure 6. When UMAP is applied to reflectance spectra (top row), the greater information content of the EMIT mosaic is manifest as a more complex topology characterized by numerous tendrils with varying degrees of differentiation from the main body of the manifold. In contrast, the multispectral sensors demonstrate diminished complexity with fewer identifiable tendrils and a more continuous structure.

The difference in manifold structure between EMIT and multispectral spectra is further accentuated when UMAP is applied to the mixture residual mosaic (bottom row). Here, spectral differences within and among tiles result in clearly separated, well-defined clusters for EMIT. The multispectral sensors are not characterized by such spectral separability. For these data, manifolds are visibly well-connected, without such clearly separable gaps. The implications of this difference in manifold structure for both discrete and continuous image analysis are discussed below.

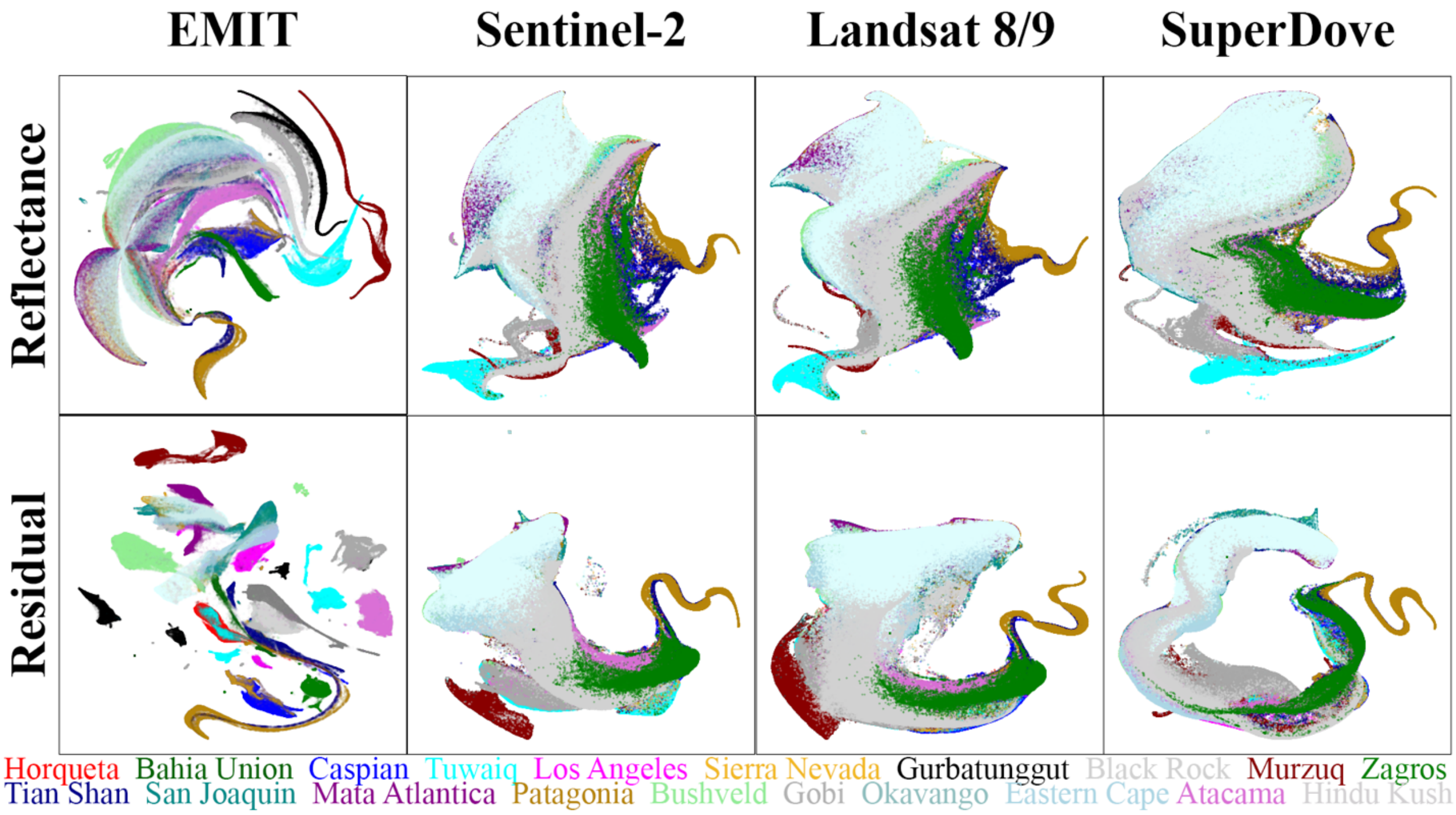

*Figure 6. Effect of spectral resolution on manifold structure. Pixels from each EMIT scene are visualized using distinct colors.*

## Joint Characterization

It has been noted previously that important information may exist at multiple scales of spectral variance in the same dataset, and that such information may be usefully examined using Joint Characterization (JC) in which bivariate distributions are used to simultaneously visualize both global and local spectral features [13]. Figure 7 illustrates JC as implemented for the EMIT reflectance mosaic. Here, the S, V, D endmembers are used as the global variance metric (x axis), and UMAP dimensions are used as the local variance metric (y axis). Tendrils at similar values of each mixture model fraction (similar x values, but distinct y values) correspond to statistically distinct clusters with broadly similar overall spectral continua but distinct absorption features (e.g., endmember variability). These tendrils frequently correspond to spatially contiguous clusters of pixels in individual EMIT scenes. This is examined in greater detail below for example EMIT scenes.

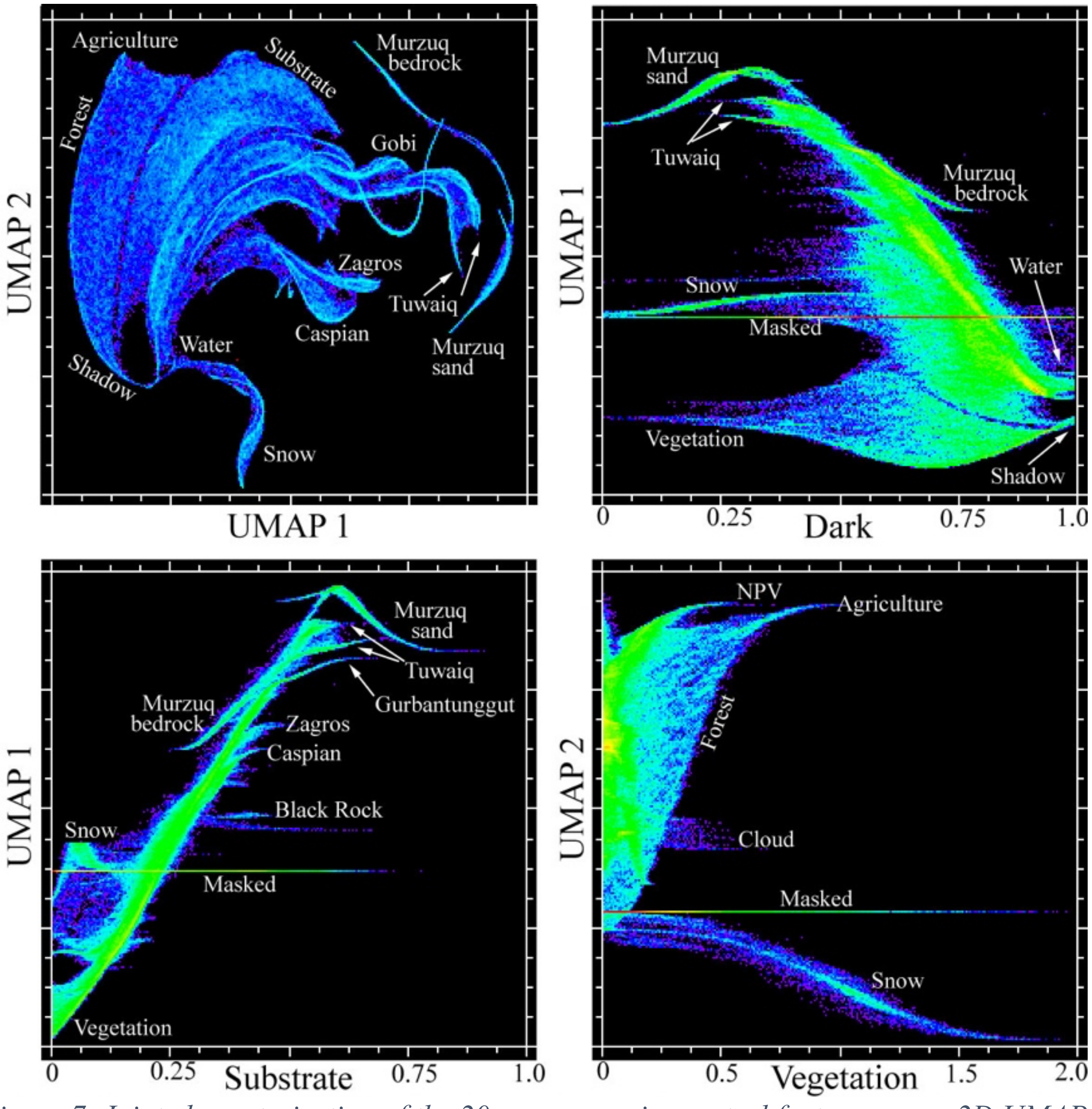

*Figure 7. Joint characterization of the 20 scene mosaic spectral feature space. 2D UMAP manifold (upper left) shows distinct 2D continua for vegetation and substrates with distinct*

*tendril continua for spectrally distinctive sands, bedrock lithology, and snow. Combining individual dimensions of 2D UMAP manifold with individual endmember fractions shows physical properties of distinct spectra. Note geographically specific lithologic endmembers in Substrate+UMAP*$_1$ *space in contrast to geographically comingled vegetation and nonphotosynthetic vegetation (NPV) endmembers in Vegetation+UMAP*$_2$ *space.*

## Single-Scene Examples

Joint characterization of individual EMIT scenes illustrates additional spectral feature space structure not apparent in the 20 scene mosaic. Figure 8 shows joint characterization as applied to vegetation spectra from the single San Joaquin scene. Red, yellow, and cyan regions of interest are identified and labeled (different colors) as clearly separable clusters from the JC scatterplot (upper right), then projected onto the SVD fraction space (upper left) for context. Average spectra from pixels in all three labeled regions of interest (bottom row) clearly correspond to photosynthetic vegetation. Differences in red edge slope, mesophyll reflectance and liquid water absorptions are present, as well as subtle differences in pigment absorption at visible wavelengths (lower right).

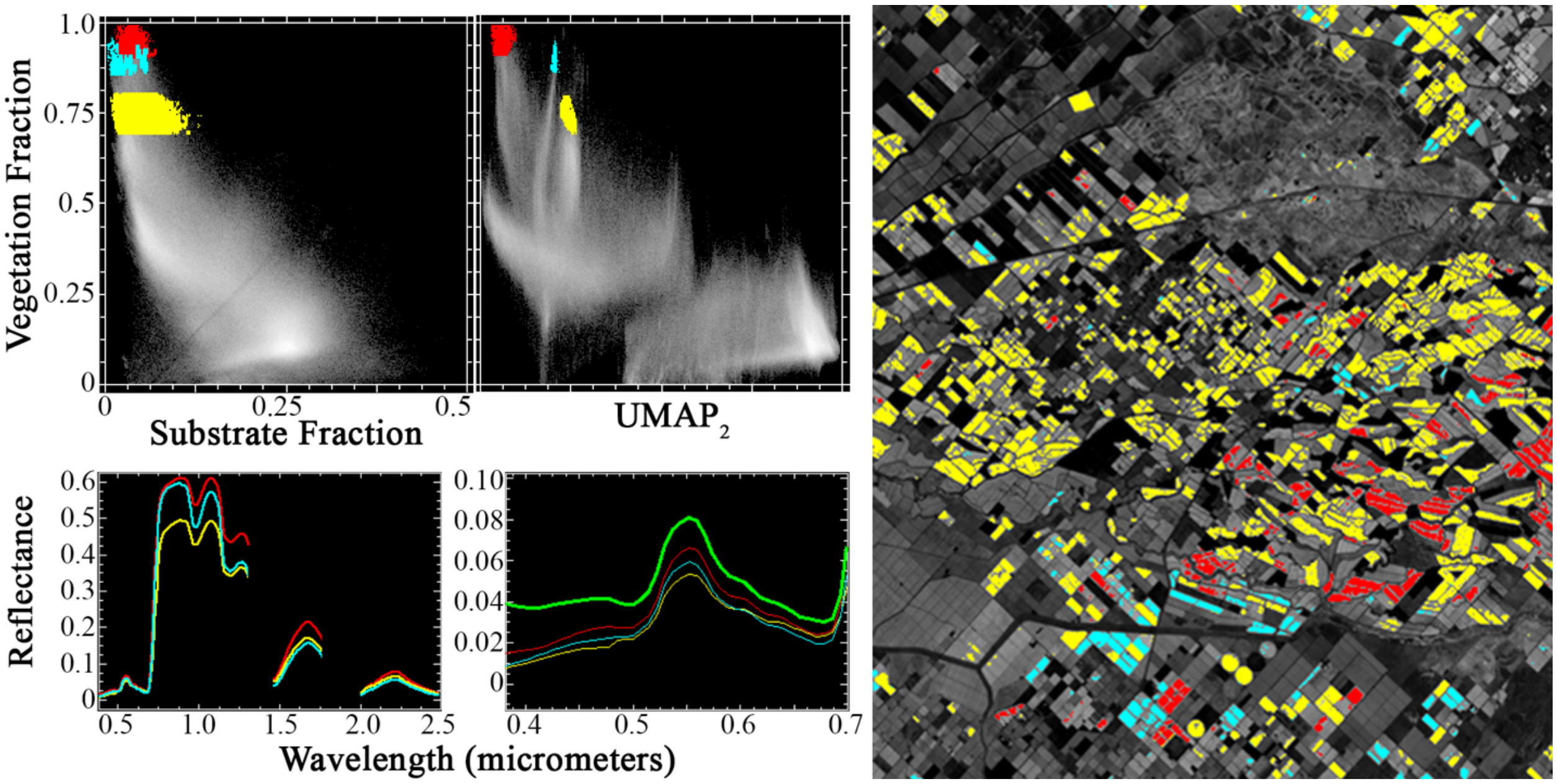

*Figure 8. Joint characterization for discrimination of vegetation spectra in the San Joaquin Valley. Red, cyan, and yellow regions of interest are clearly separable in the joint space (upper right), but not in the variance-based SVD fraction space (upper left). Mean spectra from each region of interest (bottom row) show differences in mesophyll reflectance, red edge slope, liquid water absorption, and cellulose/lignin absorption features. All regions are more absorptive throughout visible wavelengths than the mosaic V endmember (green, lower right). Clusters identified from JC are geographically coherent at the field scale in map space (right).*

Figure 9 shows joint characterization as applied to substrate spectra from the single Atacama scene. Differently colored regions of interest are identified and labeled as clearly separable clusters from the JC scatterplot (upper right), then projected onto the SVD fraction space (upper left) for context. Average spectra from pixels in all regions of interest (bottom row) clearly correspond to exposed geologic substrates. All regions are substantially more absorptive than the global sand endmember (lower left, thick red). Differences in albedo and VNIR curvature are present, as well as specific absorption features in the 2.0 to 2.5 micron region (lower right).

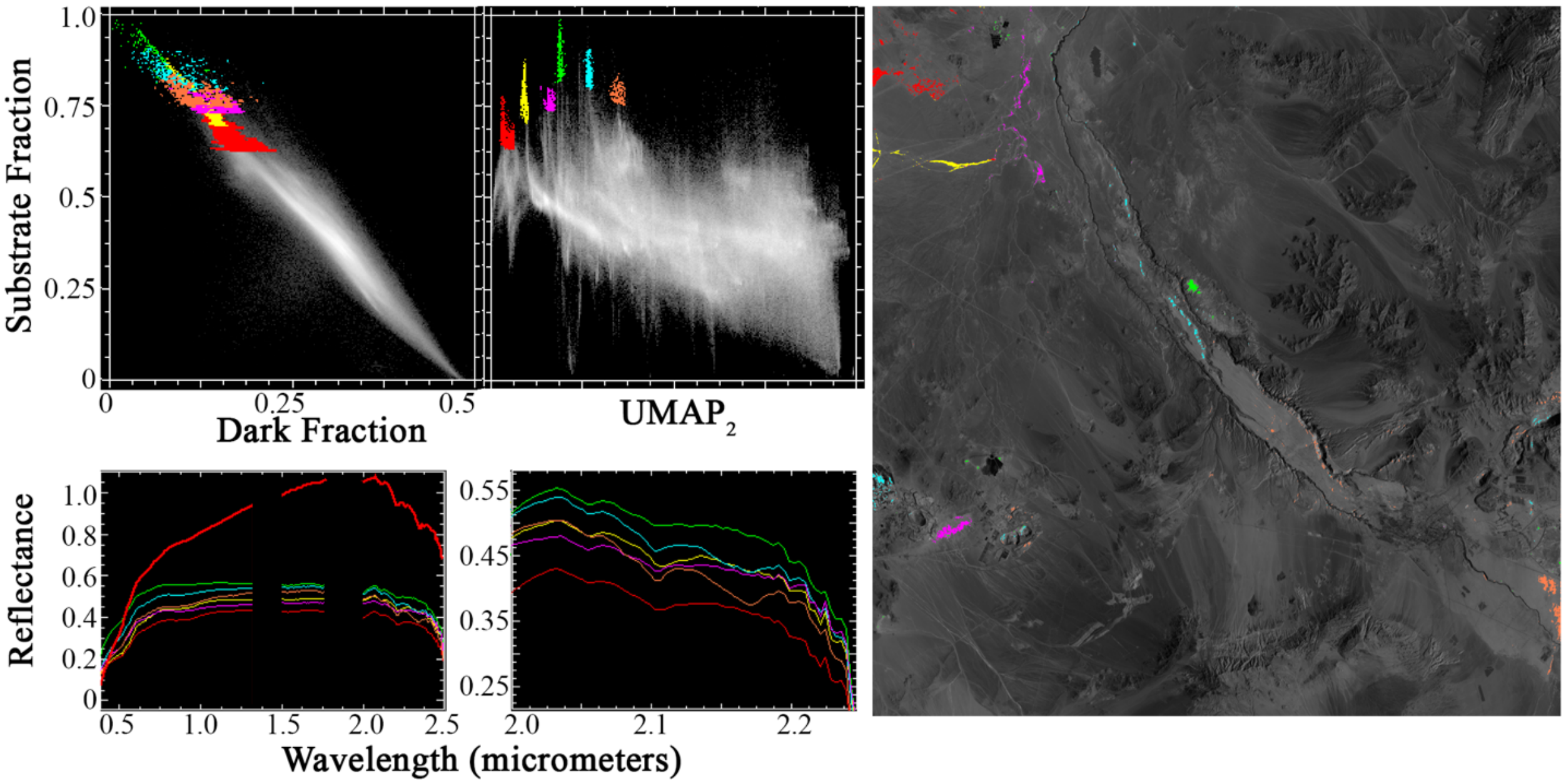

*Figure 9. Joint characterization for discrimination of substrate spectra in the Atacama Desert. Red, cyan, and yellow regions of interest are clearly separable in the joint space (upper right), but not in the variance-based SVD fraction space (upper left). Mean spectra from each region (bottom row) show differences in amplitude and curvature throughout the spectrum, including minor but perceptible differences in SWIR wavelengths (lower right). All regions are darker throughout VSWIR wavelengths than the mosaic S endmember (thick red, lower left). Clusters identified from JC are geographically coherent in map space (right).*

Figure 10 shows joint characterization as applied to dark spectra from the single Bahia Union scene. Differently colored regions of interest are identified and labeled as clearly separable clusters from the JC scatterplot (upper right), then projected onto the SVD fraction space (upper left) for context. Average spectra from pixels in all regions of interest (bottom row) clearly correspond to different shallow and suspended sediment. Differences in overall brightness and VNIR curvature are present, likely corresponding to factors like bathymetry and turbidity (lower right).

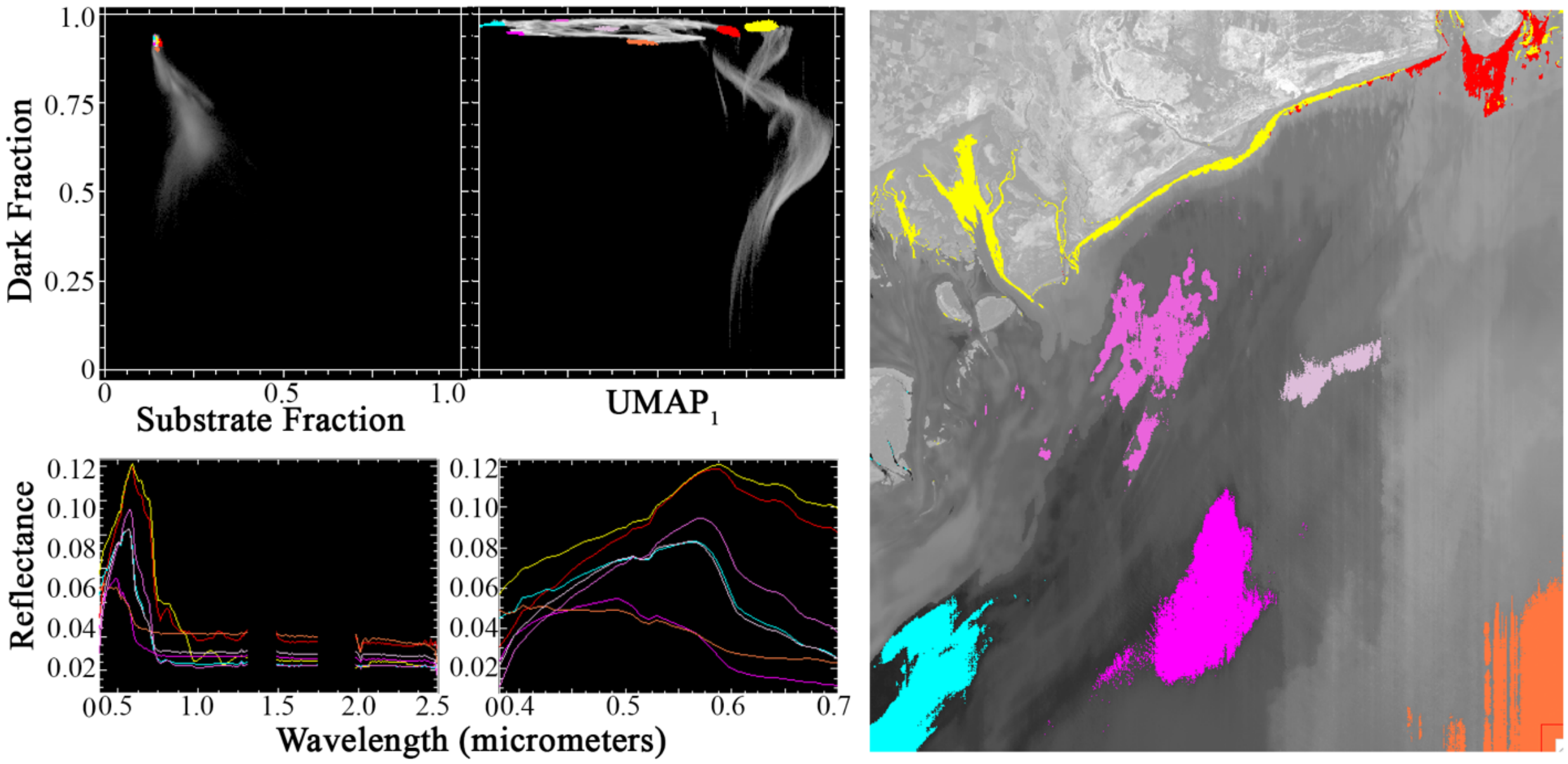

*Figure 10. Joint characterization for discrimination of dark spectra in Bahia Union coastal ocean. Regions of interest are clearly separable in the joint space (upper right), but not in the variance-based SVD fraction space (upper left). Mean spectra from each region (bottom row) show differences in amplitude and curvature throughout the spectrum, including significant differences at visible wavelengths (lower right). Clusters identified from JC are geographically coherent in map space (right).*

# Discussion

This analysis was guided by four major questions. We discuss lessons learned for each question below:

## Generality of the SVD Model

It has been acknowledged for decades that, for most of the Earth's land surface, the variance in decameter-scale multispectral satellite imagery can be contained in three dimensions. Early observations of the "brightness", "greenness", and "third" dimensions conceptualized by the Tasseled Cap [23,24] were subsequently extended to the domain of spectral mixture analysis with generalized global endmembers (EMs) corresponding to soil and rock Substrate, illuminated photosynthetic Vegetation, and Dark targets like shadow and water (S, V, and D) [16]. The SVD model has been repeatedly confirmed for larger compilations of Landsat [17,18], as well as decameter multispectral data with more spectral bands from Sentinel-2 [20]. Decameter to meter spatial scaling has been characterized using Landsat:WorldView-2 image pairs [25], as well as decameter to hectometer spatial scaling using coincident Landsat:MODIS observations [19].

Spectral unmixing was largely developed in the context of imaging spectroscopy [26–28], and models using soil, photosynthetic vegetation, and shadow have been applied to imaging spectroscopy data for decades, often with the addition of a non-photosynthetic vegetation (NPV) endmember (e.g., [26,29]). As noted above, the majority of such studies used airborne imaging spectroscopy, and so largely operated at local to regional scales. Studies of compilations of AVIRIS flight lines have also been performed [15,30–32]. Such studies largely focus on the related but distinct concept of intrinsic dimensionality, e.g., [33–35]. Those that do focus on SVD model generality are limited in spatial scope by data availability to North America, primarily California [14,15,36]. Evaluation of the generality of the SVD model with geographically and spectrally diverse EMIT data was a primary objective of this study. To our knowledge, this is the most comprehensive study to date to demonstrating generality of the SVD model for imaging spectroscopy data, and the first to do so with decameter spaceborne data.

## Feature Space Dimensionality and Topology: Hyperspectral vs Multispectral

The cross-sensor generality of the SVD model is intrinsically related to the similarity (or lack thereof) in spectral feature space dimensionality (variance partition) and topology. As noted above, intrinsic dimensionality of imaging spectroscopy data has been studied previously, but studies have been limited by both data coverage and line-to-line differences in sensor calibration and atmospheric correction. The EMIT reflectance product used for this study may exhibit substantially enhanced image-to-image radiometric stability relative to compilations of multiple flight lines from airborne sensors. The similarity in dimensionality between EMIT and simulated multispectral sensors when quantified using eigenvalues computed from reflectance – and dissimilarity when quantified using variance partition computed from the mixture residual – aligns with and extends previous results of [14,37] in clearly demonstrating that the greater spectral information content in hyperspectral image data can be effectively conceptualized as greater departure from a simple 3 endmember linear mixing model. The fundamental differences in topology between the UMAP(MR) results for EMIT versus all other sensors, discussed below, also significantly strengthen and extend this finding.

## Mixture Residual Efficacy: Hyperspectral vs Multispectral

UMAP results from EMIT spectra at full spectral resolution indicate a demonstrably distinct manifold structure from all simulated multispectral sensors (Figure 6). While this distinction is observed when examining reflectance spectra, differences are much more apparent with the mixture residual. The separability among clusters of MR spectra both within and across EMIT scenes is unambiguous. This result strongly suggests that the spectral signatures captured by EMIT can differentiate biogeophysically distinct Earth surface materials which are not resolved by multispectral sensors like Landsat and Sentinel. The further loss of the distinct tendrils observed in Landsat and Sentinel in the SuperDove MR manifolds suggests that SWIR bands are especially important for differentiation of these land cover types. This is particularly true for mineral absorptions in substrates, consistent with expectations of important information at SWIR wavelengths.

## Efficacy of Joint Characterization with EMIT Data

Figures 7-10 clearly indicate that Joint Characterization (JC) has significant potential to assist with exploratory analysis of high SNR decameter spaceborne imaging spectroscopy data. SVD mixture fractions provide natural quantities for the variance-based (global) axis of the JC. For this purpose, mixture model fractions have important advantages (e.g., physically interpretable) which are not generally true for other global metrics like PC dimensions. For the topology-based

(local) axis of the JC, UMAP scores are shown to provide useful information. Clusters identified from the JC space are consistently found to be statistically distinct and geographically coherent. Cluster position is not generally interpretable in UMAP space, but introducing S, V, D fractions effectively provides physical order.

### Limitations and Future work

While the results of this study are promising, we do note some significant limitations:

First, sampling is not truly global so spectral diversity is underrepresented. While a wide range of geologic and floristic landscapes are sampled, several important areas are not yet included. Notably: a) no scenes are included from Europe or Australia, b) only one urban area (Los Angeles) is included, c) only one major agricultural basin is sampled (San Joaquin), d) no boreal (e.g. tundra) environments are included and e) cryospheric diversity is underrepresented. Future studies with greater data coverage may significantly extend these results, particularly in the form of more extreme endmember spectra.

Second, while the generalized SVD model is effective at modeling a wide range of terrestrial environments, it is intentionally exclusive of some materials. Such materials not well-fit by the SVD model include natural materials like evaporites, cyrosphere (snow & ice), and shallow water substrates (e.g. reefs); as well as anthropogenic materials like roofing materials, plastics, and paint. While we recommend analyses to include both global and local EMs, it is likely that analyses which include significant areal coverage of evaporite pans, cryosphere, and/or urban areas may especially benefit from local EM selection and (potentially) mixture models with more than three endmembers.

Third, the results of this approach are inherently statistical, data-driven characterizations. Physical meaning, particularly of differences in cluster spectra identified from JC, does require user knowledge of reflectance spectroscopy. This approach is capable of identifying statistically distinct spectral signatures – but interpretation of the physical meaning of those features is likely to benefit from models with a different purpose. In particular, synergy with models with direct physical interpretation like Tetracorder [38] is likely to be particularly profitable.

## Conclusions

We analyze a spectrally and geographically diverse mosaic of 20 scenes from NASA's novel Earth Mineral Dust Source Investigation (EMIT) mission. We evaluate the applicability of the generalized Substrate, Vegetation, Dark (SVD) linear mixture model from previous studies to these data, and find the model to successfully fit the broad, high variance signatures in EMIT reflectance (average RMSE of non-masked pixels 3.1% ; 99% of pixels < 3.7%). EMIT data are convolved to the spectral response functions of three common multispectral sensors. We find the partition of variance of EMIT reflectance spectra to be comparable to modeled Landsat and Sentinel reflectance spectra, but significant and consistent differences to be present in partition of variance among sensors for the spectral mixture residual. Similarly, UMAP-estimated manifold structure for EMIT mixture residual is topologically distinct (more clustered) from manifold structure of the multispectral mixture residual. Joint characterization is found to effectively synergize the physical interpretability of the SVD mixture model with the statistical strengths of UMAP to effectively render additional potentially useful information. These results synthesize recent developments in hyperspectral high dimensional characterization, highlight the superb data quality from the novel EMIT mission, and demonstrate the quantitative and qualitative added value of spaceborne imaging spectroscopy over traditional multispectral satellite imaging.

## Acknowledgements

DS gratefully acknowledges funding from the USDA NIFA Sustainable Agroecosystems program (Grant # 2022-67019-36397), the NASA Land-Cover/Land Use Change program (Grant # NNH21ZDA001N-LCLUC), the NASA Remote Sensing of Water Quality program (Grant # 80NSSC22K0907), and the NSF Signals in the Soil program (Award # 2226649). CS acknowledges the support of the endowment of the Lamont Doherty Earth Observatory of Columbia University.

## Supplement

Table S1: Emit scenes used in this study. Latitude and longitude refer to the northwest corner of the scene.

| Title | Short Name | Latitude | Longitude |
|---|---|---|---|
| EMIT_L2A_RFL_001_20220909T145335_2225209_006 | Horqueta | -41.53 | -68.60 |
| EMIT_L2A_RFL_001_20220903T163129_2224611_012 | Bahia Union | -39.24 | -62.09 |
| EMIT_L2A_RFL_001_20220903T101734_2224607_026 | Eastern Cape | -33.01 | 23.50 |
| EMIT_L2A_RFL_001_20220830T065605_2224205_022 | Tuwaiq | 24.74 | 46.30 |
| EMIT_L2A_RFL_001_20220828T174405_2224012_007 | Los Angeles | 34.99 | -118.51 |
| EMIT_L2A_RFL_001_20220817T140711_2222909_021 | Murzuq | 26.30 | 12.39 |
| EMIT_L2A_RFL_001_20220815T042838_2222703_003 | Caspian | 40.12 | 54.22 |
| EMIT_L2A_RFL_001_20220815T025827_2222702_016 | Gurbantunggut | 45.68 | 88.96 |
| EMIT_L2A_RFL_001_20220814T223520_2222615_004 | Black Rock | 41.36 | -119.54 |
| EMIT_L2A_RFL_001_20220814T160517_2222611_005 | Sierra Nevada | 38.45 | -119.69 |
| EMIT_L2A_RFL_001_20220909T131308_2225208_011 | Atacama | -21.95 | -69.18 |
| EMIT_L2A_RFL_001_20220905T083937_2224806_033 | Bushveld | -24.46 | 26.61 |
| EMIT_L2A_RFL_001_20220827T043253_2223903_002 | Tian Shan | 41.95 | 77.10 |
| EMIT_L2A_RFL_001_20220814T160505_2222611_004 | San Joaquin | 37.97 | -120.41 |
| EMIT_L2A_RFL_001_20220901T034405_2224403_006 | Hindu Kush | 36.73 | 68.68 |
| EMIT_L2A_RFL_001_20220909T114035_2225207_003 | Mata Atlântica | -22.75 | -44.88 |
| EMIT_L2A_RFL_001_20220909T070044_2225204_005 | Okavango | -18.83 | 22.51 |
| EMIT_L2A_RFL_001_20220912T154138_2225510_002 | Patagonia | -49.58 | -74.14 |
| EMIT_L2A_RFL_001_20220816T070436_2222805_008 | Gobi | 41.72 | 104.40 |
| EMIT_L2A_RFL_001_20220901T052019_2224404_013 | Zagros | 27.70 | 55.64 |